# A mean-field hard-spheres model of glass

L. F. Cugliandolo

*Service de Physique de L'Etat Condensé, CEA, Orme des Merisiers, 91191, Gif–sur–Yvette Cedex, France.*

J. Kurchan, R. Monasson

*Laboratoire de Physique Theorique de l'Ecole Normale Superieure, 24 rue Lhomond, Paris, France.*

G. Parisi

*Dipartimento di Fisica, Università di Roma I, La Sapienza, Piazzale A. Moro 2, Roma, Italia*

*INFN - Sezione di Roma I*

(June, 1995)

## Abstract

We present a model of spheres moving in a high-dimensional compact space. We relate it to a mixed matrix model with a $O(N)$ invariant model plus a $P(N)$ invariant perturbation. We then study the low pressure regime by performing a diagrammatic expansion of this matrix model. Finally, we show the results from numerical simulations and we present evidence for a glassy regime at high pressures.

Glassy systems can be characterized as out-of-equilibrium systems in which the relaxation is slower than the experimental time-scale [1]. The slowness of the dynamics is due to the complexity of the energy (or free-energy) landcape, which may contain a combination of barriers, bottlenecks and even flat directions. In the case of *spin*-glasses, the complexity of the landscape is induced by the *frustration* in the spin configurations, which in turn is a consequence of the (quenched) disorder in the couplings. In structural glasses the ruggedness of the landcape does not necessarily come from a quenched disorder, but is a consequence of the geometrical constraints for the motion; the disorder is *self-induced*.

Despite this difference in the origin of disorder, both spin and structural glasses share some common features such as a long-term memory going back to the time when the system was quenched to low temperature. These 'aging' effects can be studied analytically in simple mean-field spin-glass models [2] and they are qualitatively very similar to the aging effects found in true experimental spin-glasses.

Recently there has been a considerable success in constructing the 'structural' counterpart of such simple spin-glass models, *i.e.* mean-field systems without quenched disorder having a dynamic phase transition into a glassy phase. These models include spin models [4–7], particles on a hypercubic cell [8] and a field theory with quasi-random interactions [9]. In all but the latter model, the glassy behaviour is a consequence of the discreteness of the variables: a continuous version of them behaves like a 'liquid' at all temperatures.

The purpose of this paper is to present a model of hard spheres [3] in a high dimensional (off-lattice) *compact* space. The model is related to (but different from) ordinary spheres in a flat high-dimensional space [10]: as we shall show below it has a more complicated low pressure virial expansion, but it is easier to simulate for high dimensions. We obtain analytical results for the low pressure regime, and numerical evidence for a glassy transition. We simulate the constant-pressure dynamics in the high-pressure phase, and study the ergodic properties in that case.

## I. THE MODEL

We consider a set of $N$ particles of positions $\vec{S}_a$ $(a = 1, ..., N)$ constrained to be on a $D$-dimensional sphere:

$$\frac{1}{D}\sum_{i=1}^{D}(S_i^a)^2 = 1 , \qquad \forall\, a \tag{1}$$

The 'distance' between two particles on the sphere is:

$$\frac{q^{ab}}{\sqrt{D}} = \frac{1}{D}\sum_{i=1}^{D} S_i^a S_i^b = \cos\theta^{ab} , \qquad \forall a < b \tag{2}$$

and the hard sphere condition reads:

$$-Q < q^{ab} < Q , \qquad \forall a < b \tag{3}$$

where $Q$ is the 'size' of the particles. We choose $Q$ of order one so that the angle between particles $\theta^{ab}$ on the sphere is close to $90^o$ (*cfr.* Eqs.(1),(2)). This yields a non-trivial behaviour in the high-dimensional limit with the particle number scaling with the dimension as $N = \rho D$ and the 'density' $\rho$ of order one. For every particle there is another in the antipodes of the sphere. Equivalently, we can say that we identify opposite points on the sphere. In that case, the manifold on which the particles move is no more the D-dimensional sphere but the projective plane $RP(D)$.

An alternative normalization is to impose the minimal angle between any two particles $\theta_{min}$ to be fixed and different from $90^o$ in the limit $N \to \infty$. In that case, $Q$ is of order $\sqrt{D}$ and $N$ has to grow exponentially with $D$ to have a non-trivial behaviour [3]. Spheres in a flat high-dimensional space are recovered in the limit $\theta_{min} \to 0$. We do not consider here this alternative normalization.

In order to study a constant-pressure situation, one can consider a variable radius of the sphere on which the particles move, and a coupling term between the radius of that sphere and the pressure. We shall instead follow the equivalent procedure of keeping the radius of the spherical space fixed, and considering the particle size $Q$ as a dynamic variable. The partition function reads

$$Z[\rho,P]=\int dQ\ \exp[-N^2PQ]\int\prod_{a=1}^{N}d\vec{S}_a\ \delta\left(\vec{S}_a^2-D\right)\prod_{a<b}\theta\left(Q\sqrt{D}-\left|\vec{S}_a.\vec{S}_b\right|\right)\ . \tag{4}$$

The coupling with the pressure $P$ is provided by the factor $\exp[-N^2PQ]$ in the partition function. Note that any other function of $Q$ just amounts to a (nonlinear) rescaling of the pressure.

In the Section III we shall study the response of the density of the system to a space-dependent field. In an ordinary flat space one applies a field with a strength that varies spatially as $h(t)\cos(kx)$ and then measures the corresponding space Fourier component of the density. Here we shall apply a field that pushes the particles towards the direction $(1,1,...,1)$ :

$$h(t)\sum_{ia}s_a^i \tag{5}$$

and we shall then measure the corresponding density response

$$m(t)=\frac{1}{ND}\langle\sum_{ia}s_a^i(t)\rangle\ . \tag{6}$$

Finally, let us remark that one can define a shear viscosity for this model by subjecting the system to a tangential force that varies with the 'latitude' from the equator of the spherical space, and measuring the corresponding average velocities. We do not do so here.

## II. ANALYTICAL CALCULATIONS

In the thermodynamical limit, the free-energy depends only on the pressure $P$ and on the density $\rho=N/D$. Introducing the entropy

$$S(q^{ab})=\log\int\prod_{a=1}^{N}d\vec{S}_a\ \delta\left(\vec{S}_a^2-D\right)\prod_{a<b}\delta\left(q^{ab}-\frac{1}{\sqrt{D}}|\vec{S}_a.\vec{S}_b|\right) \tag{7}$$

of all positions corresponding to a given set of overlaps $q^{ab}$ (Eq.(2)), $Z[\rho,P]$ may be rewritten as the partition function of the following matrix model

$$Z[\rho,P]=\int dQ\ \exp[-N^2PQ]\int\prod_{a<b}\left[dq^{ab}\theta(Q-|q^{ab}|)\right]e^{S(q^{ab})} \tag{8}$$

The origin of the difficulty arising in the computation of the partition function is now clearer. Whereas any rotation of the matrix $q^{ab}$ in the $N \times N$-dimensional space of the particle-overlaps leaves the entropy $S(q^{ab})$ unchanged, the hard sphere condition is only invariant under the group of permutations $P(N)$ of the particles. As a consequence, the action in Eq.(8) does not depend only upon the eigenvalues of $q^{ab}$ and the angular variables cannot be integrated out to obtain a solvable model in the large $N$ limit [14]. We shall now see that this difficulty does not prevent us from estimating the full diagrammatic expansion of the free–energy which is expected to be valid in the low $P$ regime.

### A. The diagrammatic expansion

Let us consider now a generic matrix model $M_{ab}$ whose action includes two different terms. The first one, $S_1(M)$, is $O(N)$ invariant. The second one, $\sum_{a,b} S_2(M_{ab})$, breaks the rotational symmetry but is invariant under a permutation of $P(N)$. Both parts $S_1$ and $S_2$ may be expanded in integer series of their argument. Assuming that the mean value of $M_{ab}$ is equal to zero, the first terms are quadratic in $M_{ab}$ while the following ones contain all kinds of even vertices compatible with the $O(N)$ and $P(N)$ symmetries :

$$S_1(M) = \frac{r_1}{2} \sum_{a,b} M_{ab}^2 + \frac{g_1}{N} \sum_{a,b,c,d} M_{ab} M_{bc} M_{cd} M_{da} + O(M^6) \tag{9}$$

$$S_2(M) = \frac{r_2}{2} \sum_{a,b} M_{ab}^2 + g_2 \sum_{a,b} M_{ab}^4 + O(M^6) \tag{10}$$

where the normalization factor in Eq. (9) ensures that the limit $N \to \infty$ is well defined. For simplicity we shall consider only quartic interaction terms; the extension of what follows to the general case is immediate. We therefore end up with the computation of

$$Y = \int \prod_{a,b} dM_{ab} \exp\left( -\frac{r}{2} \sum_{a,b} M_{ab}^2 - \frac{g_1}{N} \sum_{a,b,c,d} M_{ab} M_{bc} M_{cd} M_{da} - g_2 \sum_{a,b} M_{ab}^4 \right) \tag{11}$$

where $r = r_1 + r_2$. When $g_2 = 0$, $Y$ may be expanded in powers of $g_1$ and only planar diagrams survive in the large $N$ limit [14]. If we now consider the diagrams coming from the $g_2$ vertex, the self–energy will include some new contributions which may be separated in two groups :

- some diagrams are topologically equivalent to the ones arising in the $g_2 = 0$ expansion but the propagators between the $g_1$ vertices are now "dressed" with the $g_2$ interaction : such diagrams may be simply obtained by the addition of a self–energy term $r_2 + \sigma_2 \sum_{a,b} M_{ab}^2$ in $S_1(M)$, where the value of $\sigma_2$ will have to be fixed in a self-consistent way.

- the remaining "mixed" diagrams include interacting $g_1$ and $g_2$ vertices : the simplest representative of this class is

$$\frac{g_1 g_2}{N} \sum_{a,b,c,d,e,f} \langle M_{ab} M_{bc} M_{cd} M_{da} M_{ef}^4 \rangle \tag{12}$$

  where $\langle \cdot \rangle$ denotes the Gaussian bare measure over the matrix elements. Once Wick's contractions have been done in the inner loops of such diagrams, the number of free indices running inside the loops is strictly lower than the power of the $\frac{1}{N}$ factor. Following the lines of [5], it seems reasonable to conjecture that this result still holds for all "mixed" diagrams. Therefore, these diagram should not contribute in the large $N$ limit. Although we have not been able to prove this conjecture, we believe it is exact.

The $S_2$ term turns out to renormalize the self–energy of the $O(N)$ invariant model by an additional factor $\sigma_2$. As a consequence the average squared value $\langle M^2 \rangle$ equals the full propagator of the $S_1$ theory, $\langle M^2 \rangle_1 (r + \sigma_2)$ with the bare mass $r + \sigma_2$.

If we now consider $S_1(M)$ as a perturbation of the $P(N)$ invariant action $S_2$, the above reasoning will still hold and the full diagrammatic expansion is identical to the one of the $S_2$ model with a new contribution $\sigma_1$ to the self–energy. Thus, the mean squared value of the matrix element equals $\langle M^2 \rangle_2 (r + \sigma_1)$ where $\langle \cdot \rangle_2$ denotes the average value with the $S_2$ action and the bare mass $r + \sigma_1$. It is furthermore equal to the inverse of the renormalized mass. The self–consistency equations that $\sigma_1$ and $\sigma_2$ fulfill are therefore

$$\langle M^2 \rangle_1 (r + \sigma_2) = \langle M^2 \rangle_2 (r + \sigma_1) = \frac{1}{r + \sigma_1 + \sigma_2} \tag{13}$$

### B. Resummation and introduction of the disorder

We shall now see that the above diagrammatic expansion may be obtained in a compact form by introducing the new partition function (on the same lines as in Ref. [5])

$$Y[U] = \int \prod_{a,b} dM_{ab} \exp\left( -\frac{r}{2}\sum_{a,b} M_{ab}^2 - \frac{g_1}{N}\sum_{a,b,c,d} M_{ab}M_{bc}M_{cd}M_{da} - g_2 \sum_{a,b}\left(\sum_{c,d} U_{ab,cd}M_{cd}\right)^4\right) \tag{14}$$

where $U$ is a $N^2 \times N^2$ orthogonal matrix. The important point is that the annealed partition function $\overline{Y[U]}$, averaged over all orthogonal matrices $U$, has the same diagrammatic expansion as the original $Y$ (11) which corresponds to the particular choice $U_{ab,cd} = \delta_{ac}\delta_{bd}$. Let us define $T_{ab} = \sum_{c,d} U_{ab,cd}M_{cd}$ and its Lagrange multiplier $\hat{T}_{ab}$ to rewrite

$$\begin{aligned} Y[U] = \int \prod_{a,b} dM_{ab}dT_{ab}d\hat{T}_{ab} \exp\Bigg( &-\frac{r}{2}\sum_{a,b} M_{ab}^2 - \frac{g_1}{N}\sum_{a,b,c,d} M_{ab}M_{bc}M_{cd}M_{da} \\ &- g_2\sum_{a,b} T_{ab}^4 + \sum_{a,b}\hat{T}_{ab}T_{ab} - \sum_{a,b,c,d}\hat{T}_{ab}U_{ab,cd}M_{cd}\Bigg) \end{aligned} \tag{15}$$

If we now average $Y[U]$ over all possible choices of the orthogonal matrix $U$, we obtain [5]

$$\begin{aligned} \overline{Y[U]} = \int \prod_{a,b} dM_{ab}dT_{ab}d\hat{T}_{ab} \exp\Bigg( &-\frac{r}{2}\sum_{a,b} M_{ab}^2 - \frac{g_1}{N}\sum_{a,b,c,d} M_{ab}M_{bc}M_{cd}M_{da} \\ &- g_2\sum_{a,b} T_{ab}^4 + \sum_{a,b}\hat{T}_{ab}T_{ab} + G\left(\sum_{a,b}\hat{T}_{ab}^2\sum_{c,d}M_{cd}^2\right)\Bigg) \end{aligned} \tag{16}$$

where

$$G(t) = \log\int_{-1}^{1} dx(1-x^2)^{N/2}e^{iNx\sqrt{t}} \tag{17}$$

Introducing the order parameters $\frac{1}{N^2}\sum_{a,b}\hat{T}_{ab}^2$ and $\lambda = \frac{1}{N^2}\sum_{a,b}M_{ab}^2$ and their respective Lagrange multipliers $\tau$ and $\lambda$, we see from the above expression (16) for $\overline{Y[U]}$ that the integral over the $\hat{T}$ variables is purely Gaussian and may be analytically performed. The $M$ and $T$ variables, whose vertices are respectively $O(N)$ and $P(N)$ invariant, are thus coupled only through an additive renormalization of their masses by $\lambda$ and $\tau$ respectively. Therefore, the introduction of the disorder and the annealed average we have carried out has permitted us to find another partition function which must be equal to the whole resummation of the diagrammatic expansion of $Y$ exposed in the previous paragraph, provided that Eq.(13) is fulfilled, that is

$$\langle M^2\rangle(\lambda) = \langle T^2\rangle(\tau) = \frac{1}{\tau+\lambda} \tag{18}$$

It is possible that the disordered model also yields information about the high pressure phase.

### C. The hard spheres free–energy

Let us now apply the previous results to our original problem of interacting hard spheres. Starting from Eq.(8), we introduce the orthogonal matrix $U$

$$Z[U,\rho,P] = \int dQ\ \exp[-N^2PQ] \int \prod_{a=1}^{N} d\vec{S}_a \delta\left(\vec{S}_a^2 - D\right) \prod_{a<b} \theta\left(Q\sqrt{D} - \left|\sum_{c<d} U_{ab,cd}\vec{S}_c.\vec{S}_d\right|\right) \tag{19}$$

and average over all possible choices of such matrices to find

$$\overline{Z[U,\rho,Q]} = \int dQ d\lambda d\tau e^{-N^2F(\lambda,\tau,Q,\rho,P)} \tag{20}$$

where

$$F(\lambda,\tau,Q,\rho,P) = \lambda F_p(\lambda,\rho) + \frac{\tau}{2}F_s(\tau,Q) - \frac{1}{4}\log(\tau+\lambda) + QP \tag{21}$$

and the two free-energies corresponding to the $O(N\times N)$ and $P(N)$ interactions respectively equal

$$F_p(\lambda,\rho) = \lim_{N,D\to\infty} -\frac{1}{\lambda N^2}\log\int\prod_{a=1}^{N} d\vec{S}_a\delta\left(\vec{S}_a^2 - D\right)\exp\left(-\frac{\lambda}{4D}\sum_{a\neq b}(\vec{S}_a.\vec{S}_b)^2\right) \tag{22}$$

at fixed density $\rho = \frac{N}{D}$ and

$$F_s(\tau,Q) = -\frac{1}{\tau}\log\int_{-Q}^{Q}\frac{dx}{\sqrt{2\pi}}\exp\left(-\frac{\tau}{2}x^2\right) \tag{23}$$

The saddle-point equations with respect to $\tau$ and $\lambda$ read

$$\left\langle \frac{(\vec{S}_a.\vec{S}_b)^2}{D} \right\rangle_p = \langle x^2 \rangle_s = \frac{1}{\tau + \lambda} \tag{24}$$

where the measures $\langle . \rangle_p$ and $\langle . \rangle_s$ correspond to the free-energies (22) and (23) respectively. Defining explicitly the latter,

$$\mathcal{P}(x) = \frac{\exp\left(-\frac{\tau}{2}x^2\right)}{\int_{-Q}^{Q} dy \exp\left(-\frac{\tau}{2}y^2\right)} \tag{25}$$

the saddle-point over $Q$ reads

$$\mathcal{P}(Q) = P \tag{26}$$

Interpreting $\mathcal{P}(x)$ as the probability distribution of the overlap between two spheres, identity (26) expresses the usual relation between the pressure and the value of the pair correlation function at contact in hard spheres systems [11]. Note that the low pressure theory therefore predicts a (troncated) Gaussian distribution of the overlaps between the spheres while the sign of $\tau$ appearing in (25) may become negative at sufficiently high pressures. This happens when many spheres are touching themselves. The probability $\mathcal{P}$ of their mutual overlap being equal to $Q$ (or $-Q$) it then becomes higher than the probability $\mathcal{P}(x = 0)$ of finding "orthogonal" spheres.

At low $Q$, the pressure of this solution diverges as

$$P = \frac{A}{Q - Q_{min}} \tag{27}$$

where $A =$ and $Q_{min} = \sqrt{1 - \frac{1}{\rho}}$.

## III. NUMERICAL SIMULATIONS - THE GLASS TRANSITION

In order to locate a possible transition, we have performed simulations with Monte Carlo dynamics. The simulation starts from a configuration without overlaps between particles, and a large value of $Q$ (corresponding to small particles or, equivalently, low density). The particles are moven randomly, and the changes are accepted if they do not violate the hard-sphere condition.

In a first set of simulations (corresponding to infinite pressure in the transient), every $\Delta t$ steps an attempt is made to enlarge all the particles by an amount $\delta Q$: $Q \to Q - \delta Q$. If the change in size does not generate an overlap, then it is accepted. When a given target value $Q_{target}$ is reached, the system evolves at fixed $Q = Q_{target}$. Different annealing speeds correspond, in this set of simulations, to smaller proposed changes in $Q_{target}$ and larger intervals between changes of $Q_{target}$.

A second set of simulations was done at constant pressure as follows: the particles are moven at random as before, never accepting overlaps between them. After every motion an attempt is made to change the particle size. If the change generates an overlap, then it is rejected. If it does not generate an overlap, then it is accepted with the Monte Carlo probability associated to an energy $N^2 \; P \; Q$ (*cfr.* Eq (4)). At not too high densities, we have observed that the number of sweeps needed to thermalize roughly scales as $N^2$.

We have done simulations for $\rho = 2$ and $D = 4$, 8, 16, 32 and 64. For dimensions up to 16, the system goes at high densities to a regular structure (see below). We have done longer simulations for $D = 32$ and 64. Fig. 1 shows for $D = 64$ the plot of average interparticle 'squared distance' $\sum_{a \neq b} q_{ab}^2 / N(N-1)$ versus particle size $Q$ obtained with the infinite-pressure algorithm for four different annealing times (each one half of the previous, the longest run consisting of about $10^7$ sweeps), together with the analytical curve. One can see the departure of the numerical points from the low-density solution at a value of $Q < 1$. The dependence on the annealing velocity becomes more important at higher densities. However, we have checked that the correction due to the finite annealing time vanishes as the annealing time to a power close to $-0.5$. The limit of infinitely slow annealing yields a value that is different at high densities from the analytic computation for the liquid phase.

More detailed information is obtained by plotting the histogram of normalized relative distances $q_{ab}/Q$ for different particle sizes $Q$ (Fig. 2). As the particle size is increased the tipical distances become more and more concentrated near the minimal distance $Q$, *i.e.* most spheres are almost touching.

The low-density solution of Section II predicts a Gaussian form for this distribution in the low-density phase. By comparing this prediction with the numerical data we can see at which density the low-density solution breaks down.

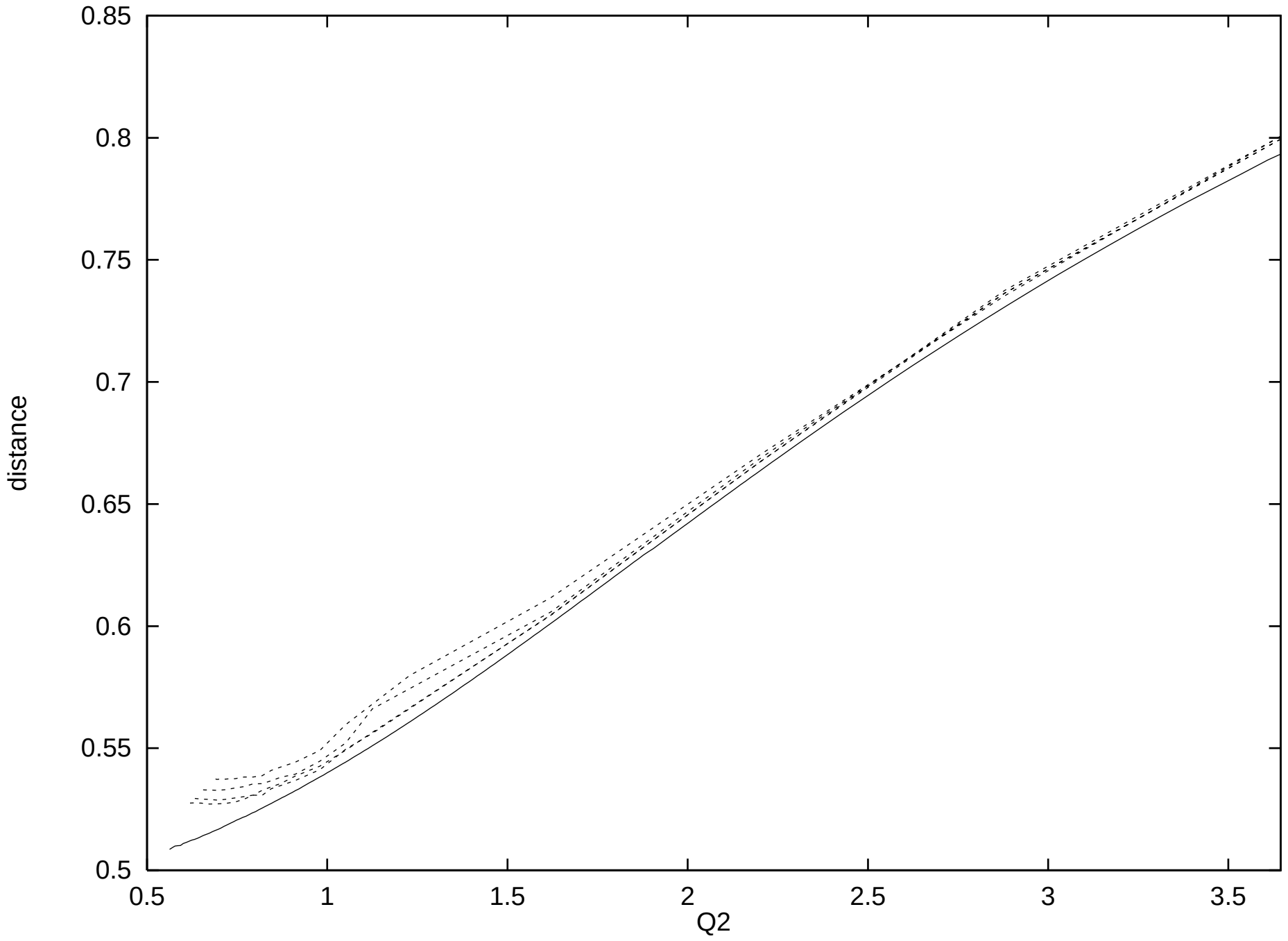

FIG. 1. Mean squared distance $\sum_{a\neq b} q_{ab}^2/N(N-1)$ between particles vs. particle size $Q^2$ for D=64, N=128. Four different annealing times.

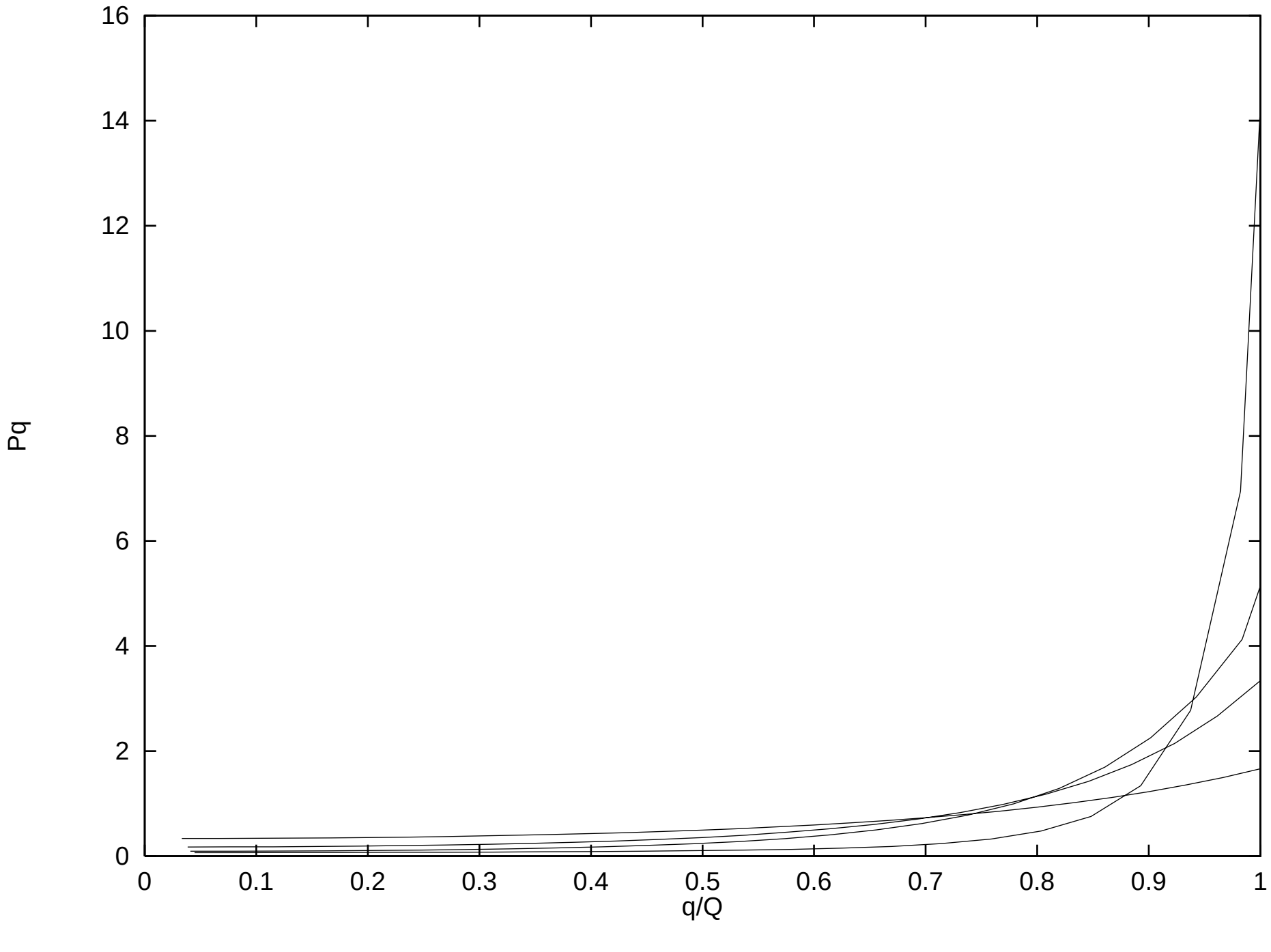

FIG. 2. Histogram of $q_{ab}$ vs. $q/Q$ for $Q$ of $0.791960, 0.862670, 0.919239$ and $1.06066$

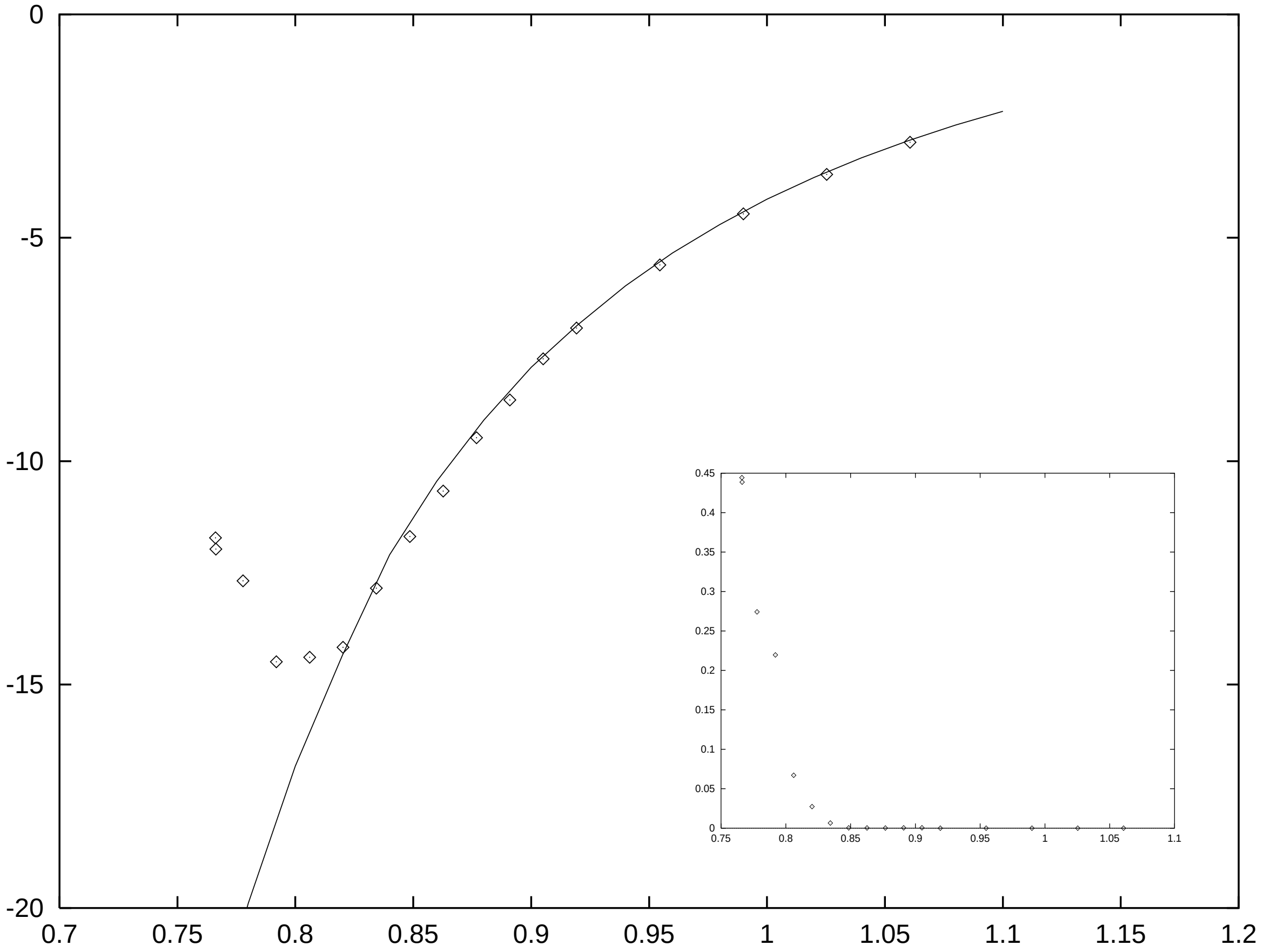

FIG. 3. $\hat{\tau}$ vs $Q$ (full line) analytical result for the low-density phase. Coefficient in the Gaussian fit of the histogram of $q_{ab}$ from the simulation. In the inset the error corresponding to the Gaussian fit.

In figure 3 we plot the coefficient of the Gaussian obtained by doing a numerical fit of the data versus the analytical prediction of the previous section. The agreement is good up to a value of $Q$ around 0.84. A further confirmation of the transition density is obtained by computing the error in the Gaussian fit. The inset of Fig. 3 shows that the error is negligible for $Q > 0.84$ and starts growing for smaller values of $Q$ (corresponding to the high-density phase). This means that, for $Q < 0.84$, the distribution $P(q^{ab})$ is definitely not a Gaussian.

With the constant pressure algorithm we have performed a slow annealing in pressure. The result of $Q$ vs. $P$ for $D = 32, \rho = 2$ (the volume versus pressure curve for this model) is shown in Fig. 4. Again, the analytical low-pressure curve breaks away from the numerical one at a value of $Q$ consistent with the previous one. In that figure we also show the pressure as obtained from the probabilty of $q_{ab}$ at $q_{ab} = Q$ (see Ref. [11]). An asymptotic form of the pressure (27) describes well the data for the longest run at low $Q$ and $D = 64$, the values of $A$ and $Q_{min}$ being 0.4 and 0.76 respectively. These values are close but different from the theoretical low density predictions of $A = *$ and $Q_{min} = \frac{1}{\sqrt{2}}$.

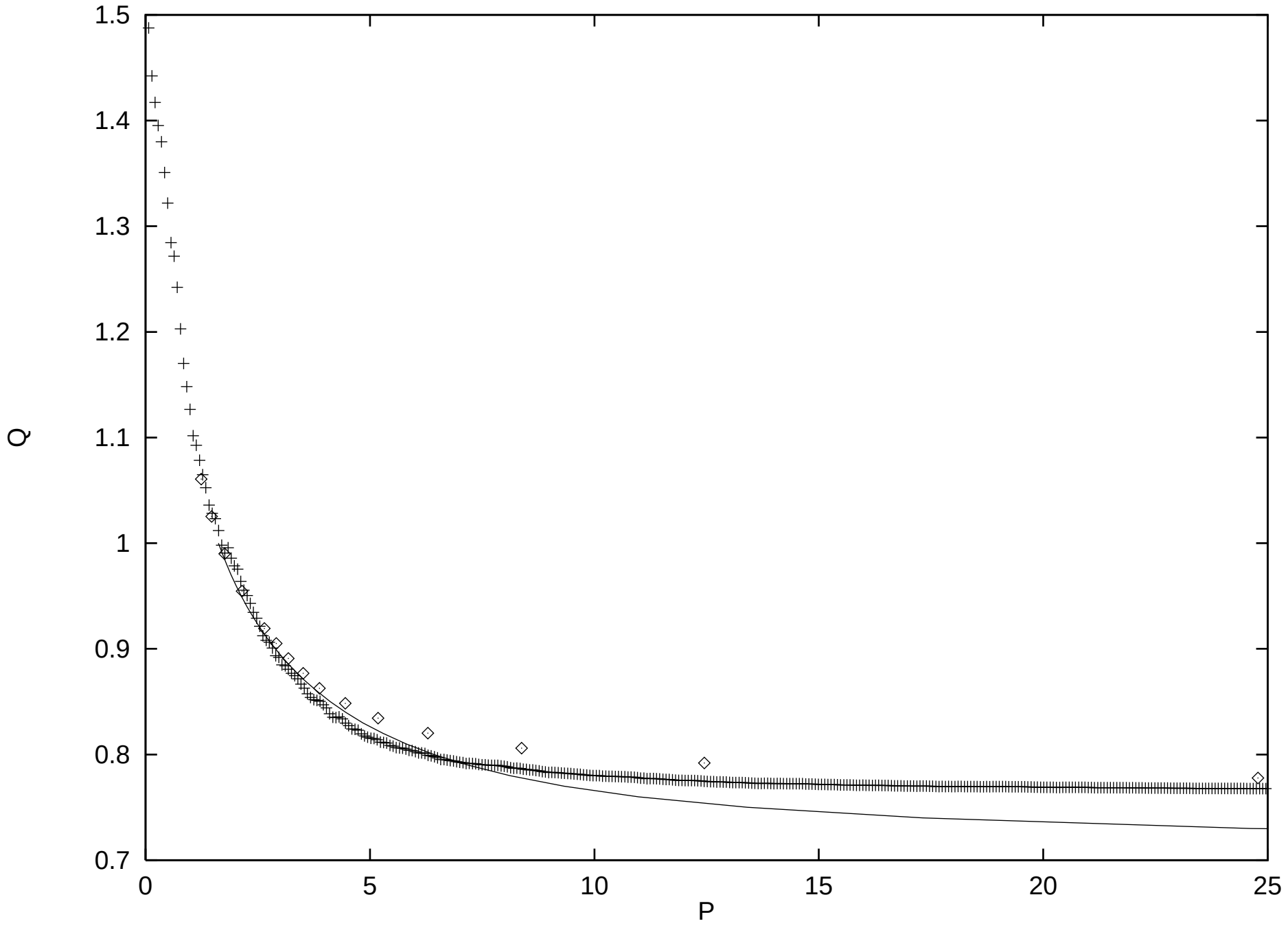

FIG. 4. Sphere diameter Q as a function of the pressure p. Diamonds : pressure estimated from the histogram $P(q^{ab})$ for $D = 64$ using the longest run. Crosses : annealing in the pressure for $D = 32$. Full line : low-density analytical calculation.

From these simulations we see that there is some sort of dynamical freezing at large pressures. In order to try to undestand the nature of the dynamical high-pressure regime, we have performed a numerical 'aging experiment' at $P = 8.4$ $(D = 32, \rho = 2)$. We perform a rapid 'quench' in pressure down to $P = 8.4$ in the presence of a constant field $(h = 30)$ * as in equation (5). The conjugate density Eq. (6) tends to stabilize after a transient. We now cut the field at different times $t_w$, and observe the decay of the density response (6) for further times $t$. The results are shown in Fig. 5, for three different $t_w$, together with (in the inset) the evolution of the density reponse for constant field.

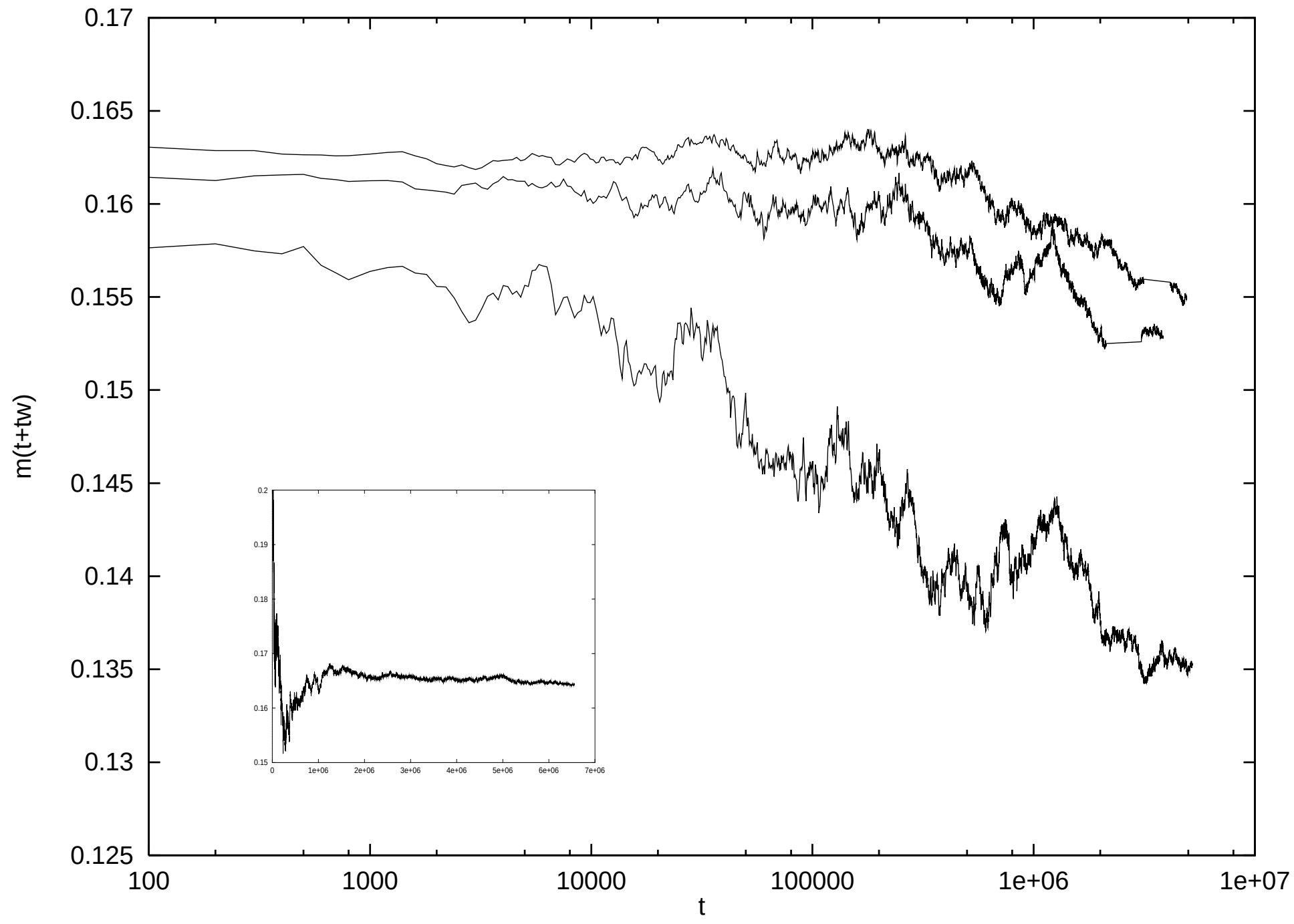

FIG. 5. Density response in the aging simulation $m(t+t_w)$ vs. $t$, log-linear plot. The waiting times are $t_w = 25.10^4$, $63.10^4$ and $102.10^4$. Inset : density response for constant field; the vertical range is from 0.15 up to 0.20. The time axis is the same for both graphs.

We see that the remanent density response has a very slow decay after the field has been cut off. The spheres are very blocked in a position that was generated under the field, and have difficulty in rearranging themselves. The system in which the field has been cut off first has the fastest decay of the density response. The question now is whether the system has 'true' or 'weak' ergodicity breaking, as we shall discuss below.

## IV. DISCUSSION

Let us now discuss the physics of this model, in view of the results we have obtained and of what we know from spin-glass dynamics. First of all we must discuss possible finite-size effects. A finite system cannot have a sharp transition; at finite pressure it will eventually equilibrate, and thus cannot have a glassy regime either. We have simulated the system with $D = 16, \rho = 2$ and found evidence of equilibration: the distribution of overlaps tends to have a few peaks, suggesting a fall into a 'crystalline' state. Though we have not performed a detailed analysis of the dependence of the equilibriation time upon $D$ and $N$, we know that it rapidly increases with both parameters. For the larger sizes we have presented (*e.g.* $D = 32, 64,\ \rho = 2$) we have observed no evidence for the formation of a crystalline state, the resulting function $P(q^{ab})$ being smooth.

Let us then discuss what is the behaviour of the model for large but finite times (as compared to the system size). The first question that arises is what is the role of the annealing procedure or, rather, what would it be in a system with $N \to \infty$. In order to distinguish the dependence upon different annealing velocities from aging effects, we shall consider first the asymptotic value of *one-time* quantities (distribution of interparticle distances, density, etc). Let us be more precise and consider the following procedure: we start with the system at low-density and we increase the pressure with a given velocity up to a given final pressure. After that we let the system relax at constant pressure for a long subsequent time. The question is now whether the asymptotic value of, say, the distribution of interparticle distances depends upon the annealing velocity. Our data are compatible with the assumption that there is no such dependence. In this respect this model behaves as real spin-glasses which are different from ordinary glasses in that the large time limit of one-time quantities are independent of the annealing procedure [15].

In the high-pressure regime there is obviously some form of slow dynamics (see Fig. 5). The question now arises as to whether we are in the presence of 'strong' or (as is the case in spin glasses) 'weak' ergodicity breaking. In the first case, the system remains confined in a small region of phase-space and remembers forever the conditions in which it was prepared. In the 'weak' ergodicity breaking scenario, there is no confinement within a small region of phase space. For generic initial conditions after a sufficiently large time the system forgets any perturbation that lasted for a finite time (and in particular the initial condition). However, as the system ages it moves slower and slower, and the time needed to forget any perturbation grows with the time during which the perturbation acted.

In the context of our 'aging experiment' of the Section III, the questions are whether the density response decays to zero for any $t_w$ (in which case the ergodicity-breaking is 'weak') and whether the typical decay time depends on $t_w$ ('aging').

The results shown in Fig. 5 seem to point in the direction of weak ergodicity with aging, but they are hardly conclusive. The decay is extremely slow, an effect that has already been observed in the models with self-induced disorder that appeared recently in the litterature. For the related matrix model of Ref. [8] it is clear that the the auto-correlation function decays to zero but it does so extremely slowly. For $t_w \gg 1$ and $t_w/\tau + t_w = O(1)$ : $C(\tau + t_w, t_w) \sim (t_w/\tau + t_w)^{0.05}$.

In this note, we have presented a spherical mean-field model for hard-spheres whose dynamics is potentially solvable in the high pressure phase. We believe that it can provide an analytical description of such typical glassy phenomena as the viscosity increase around the glassy transition and aging effects [16] but possibly not remanent dependences on the annealing procedures.

ACKNOWLEDGEMENTS L. F. C. acknowledges support from the the EU HCM grant ERB4001GT933731.